\documentclass[10.5pt,a4paper,oneside]{extarticle}
\usepackage[T1]{fontenc} % package for character coding
\usepackage[left=25mm,right=20.9mm,top=4.25cm,bottom=3.75cm,headsep=0.25cm,footskip=0.25cm]{geometry}

\usepackage[english]{babel}
    \usepackage[affil-it]{authblk}
    \usepackage{etoolbox}
    \usepackage{lmodern}
    \usepackage{setspace}
    \usepackage{amssymb,amsthm,amsmath}
    \usepackage{tikz}
    \usepackage{listings}
    \usepackage{color}
    \usepackage{caption}
    \usepackage{placeins}
    \usepackage{booktabs}
    \usepackage{enumerate}
    \usepackage{graphicx}
    \usepackage{titlesec}
    \usepackage{mathptmx}
    \usepackage{anyfontsize}
    \usepackage{t1enc}
    \usepackage[utf8]{inputenc}
    \usepackage{fancyhdr}
    \usepackage{float,xspace}
    \usepackage{scalefnt}
	
\usepackage{soul} % to be able to use \hl{} command
\usepackage{tabulary}
\usepackage[colorlinks=true,linkcolor=blue]{hyperref} % added by AO

\usepackage{textcase}

\usepackage[colorinlistoftodos]{todonotes}

\usepackage{bm,subfig}

\DeclareMathOperator*{\minimize}{minimize}

\titlespacing*{\section}
{0pt}{12pt}{0pt}
\titlespacing*{\subsection}
{0pt}{12pt}{0pt}
\titlespacing*{\subsubsection}
{0pt}{12pt}{0pt}
\titleformat{\section}
  {\huge\sffamily\Large\bfseries\color{black}}
  {\thesection}{1em}{}
    \titleformat{\subsection}[block]{\bfseries}{\thesubsection}{.5em}{}
    \titleformat{\subsubsection}[block]{\bfseries}{\thesubsubsection}{.5em}{}

\titleformat{\section}{\fontsize{12}{19}\bfseries}{\thesection}{1em}{}

\usepackage{mathptmx} % 'Times new roman'

\makeatletter

\patchcmd{\@maketitle}{\LARGE \@title}{\fontsize{14}{19.2}\selectfont\@title}{}{} % original: 18, 19.2
\makeatother
\title
{
	\vspace{-5cm}
	\begin{minipage}{\textwidth}	
	\hspace{-20pt}\vspace{100pt}%\includegraphics[width=9.5cm]{ICA_logo.jpeg}
	\hspace{-65pt}%\includegraphics[width=10.2cm]{ICA_text2022.png}
	\end{minipage}
	\textbf{Weighted pressure matching based on kernel interpolation\\for sound field reproduction}
	\author[ ]{Shoichi KOYAMA and Kazuyuki ARIKAWA$^{(1)}$}
  	\affil[(1)]{Graduate School of Information Science and Technology, The University of Tokyo, Japan, koyama.shoichi@ieee.org}
}
\date{}

\begin{document}

% \begin{titlepage}
\clearpage
\setcounter{page}{1}
\maketitle
\thispagestyle{empty}
\fancypagestyle{empty}
{	
	\fancyhf{} \fancyfoot[R]
	{
		\vspace{-2cm}%\includegraphics[width=17cm]{ICA_bar2022.png}
	}
}

\subsection*{\fontsize{10.5}{19.2}\uppercase{\textbf{Abstract}}}
{\fontsize{10.5}{60}\selectfont A sound field reproduction method called weighted pressure matching is proposed. Sound field reproduction is aimed at synthesizing the desired sound field using multiple loudspeakers inside a target region. Optimization-based methods are derived from the minimization of errors between synthesized and desired sound fields, which enable the use of an arbitrary array geometry in contrast with integral-equation-based methods. Pressure matching is widely used in the optimization-based sound field reproduction methods because of its simplicity of implementation. Its cost function is defined as the synthesis errors at multiple control points inside the target region; then, the driving signals of the loudspeakers are obtained by solving a least-squares problem. However, in pressure matching, the region between the control points is not taken into consideration. We define the cost function as the regional integration of the synthesis error over the target region. On the basis of the kernel interpolation of the sound field, this cost function is represented as the weighted square error of the synthesized pressures at the control points. Experimental results indicate that the proposed weighted pressure matching outperforms conventional pressure matching.}

\noindent{\\ \fontsize{11}{60}\selectfont Keywords: Sound field reproduction, Kernel interpolation, Weighted pressure matching} % at least 1 keyword is required (maximum of 5 keywords)

\fontdimen2\font=4pt

\section{\uppercase{Introduction}}
% \section{Introduction}

Sound field reproduction is aimed at synthesizing spatial sound using multiple loudspeakers (or secondary sources). Such a technique can be applied to virtual/augmented reality audio, generation of multiple sound zones for personal audio, and noise cancellation in a spatial region. 

Sound field reproduction methods can be classified into two major categories: \textit{integral-equation-based} and \textit{optimization-based methods}. The integral-equation-based methods are developed from the boundary integral representations derived from the Helmholtz equation, such as \textit{wave field synthesis} and \textit{higher-order ambisonics}~\cite{Berkhout:JASA_J_1993,Spors:AES124conv,Poletti:J_AES_2005,Ahrens:Acustica2008,Wu:IEEE_J_ASLP2009,Koyama:IEEE_J_ASLP2013}. The optimization-based methods are derived from the minimization of a certain cost function defined for synthesized and desired sound fields in a target region, such as \textit{pressure matching} and \textit{mode matching}~\cite{Nelson:J_SV_1993,Kirkeby:JASA_J_1996,Daniel:AES114conv,Poletti:J_AES_2005,Betlehem:JASA_J_2005,Ueno:IEEE_ACM_J_ASLP2019,Koyama:IEEE_ACM_J_ASLP2020}. Many integral-equation-based methods require the array geometry of loudspeakers to have a simple shape, such as a sphere, plane, circle, or line, and driving signals are obtained from a discrete approximation of an integral equation. In optimization-based methods, the loudspeakers can be placed arbitrarily, and driving signals are generally derived as a closed-form least-squares solution. In particular, pressure matching is widely used among the optimization-based methods because of its simplicity of implementation. Pressure matching is based on synthesizing the desired pressures at a discrete set of control points placed over the target region. 

An issue of pressure matching is that the region between the control points is not taken into consideration because of the discrete approximation. Therefore, its reproduction accuracy can deteriorate when the distribution of the control points is not sufficiently dense. However, the smaller the number of control points, the better in practice, because the transfer functions between the loudspeakers and control points are generally measured in advance. We propose an optimization-based sound field reproduction method called \textit{weighted pressure matching}. We define the cost function as the regional integration of the synthesis error over the target region. On the basis of the kernel interpolation of the sound field~\cite{Ueno:IEEE_SPL2018,Ueno:IEEE_J_SP_2021}, this cost function is represented by the pressures at the control points with the regional integration of kernel functions. When the same kernel functions are used for interpolating primary and secondary sound fields, i.e., the desired field and the sound field of each loudspeaker, respectively, the driving signal is obtained as the solution of weighted least squares problem with the weighting matrix defined by using the kernel functions. Experimental evaluation comparing pressure matching and weighted pressure matching is performed. 

\section{\uppercase{Problem statement and prior works}}

\subsection{Problem formulation}

\begin{figure}
\centering
\includegraphics[width=6cm]{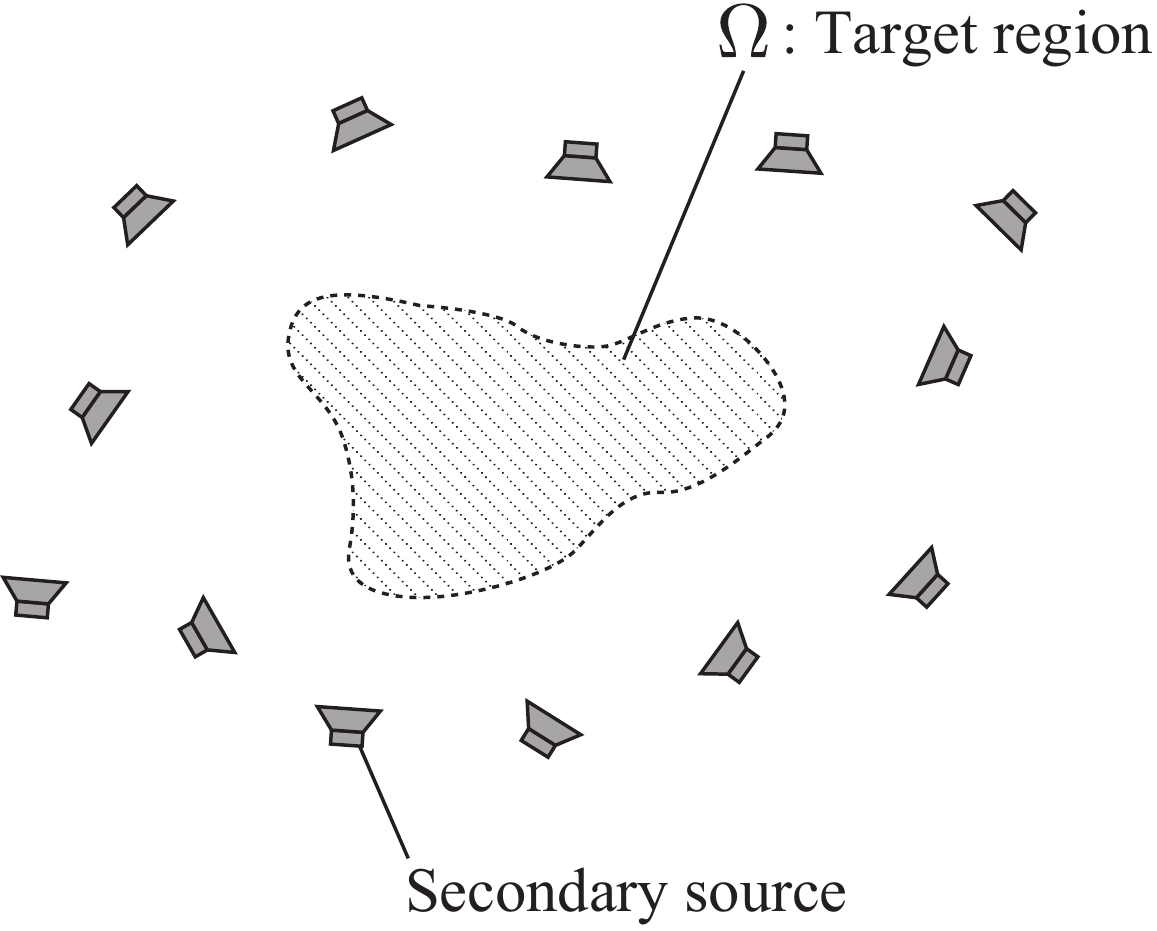}
\caption{The desired sound field is synthesized in the target region $\Omega$ using multiple secondary sources.}
\label{fig:sfc}
\end{figure}

Suppose that $L$ secondary sources (loudspeakers) are placed around a target region $\Omega \subset \mathbb{R}^3$ as shown in Fig.~\ref{fig:sfc}. The sound field $u_{\mathrm{syn}}(\bm{r}, \omega)$ at the position $\bm{r}\in\mathbb{R}^3$ and angular frequency $\omega\in\mathbb{R}$ synthesized using the secondary sources is represented as
\begin{align}
 u_{\mathrm{syn}} (\bm{r},\omega) = \sum_{l=1}^L d_l(\omega) g_l(\bm{r},\omega),
\end{align}
where $d_l(\omega)$ is the driving signal of the $l$th secondary source ($l\in\{1,\ldots,L\}$), and $g_l(\bm{r},\omega)$ is the transfer function from the $l$th secondary source to the position $\bm{r}$. The transfer functions $g_l(\bm{r},\omega)$ are assumed to be known by measuring or modeling them in advance. Hereafter, the angular frequency $\omega$ is omitted for notational simplicity. 

The goal of sound field reproduction is to obtain $\{d_l\}_{l=1}^L$ of the $L$ secondary sources so that $u_{\mathrm{syn}}(\bm{r})$ coincides with the desired sound field, denoted by $u_{\mathrm{des}}(\bm{r})$, inside $\Omega$. We define the cost function to determine the driving signal $\{d_l\}_{l=1}^L$ as
\begin{align}
 J &= \int_{\Omega} \left| \sum_{l=1}^L d_l g_l(\bm{r}) - u_{\mathrm{des}}(\bm{r}) \right|^2 \mathrm{d}\bm{r}. %\notag\\
 %&= \int_{\Omega} \left| \bm{g}(\bm{r})^{\mathsf{T}}\bm{d} - u_{\mathrm{des}}(\bm{r}) \right|^2 \mathrm{d}\bm{r},
\label{eq:cost}
\end{align}
%where $\bm{g}(\bm{r})=[g_1(\bm{r}), \ldots, g_L(\bm{r})]^{\mathsf{T}} \in \mathbb{C}^L$ and $\bm{d} = [d_1, \ldots, d_L]^{\mathsf{T}} \in \mathbb{C}^L$ are the vectors of the transfer functions and driving signals, respectively. The optimal driving signal $\bm{d}$ can be obtained by solving the minimization problem of $J$. 
The optimal driving signal can be obtained by solving the minimization problem of $J$. 

\subsection{Pressure matching}

Since it is difficult to solve the minimization problem of $J$ owing to the regional integration over $\Omega$, several methods based on the approximation of $J$ have been proposed. A simple strategy for solving it is to discretize the target region $\Omega$ into multiple control points, which is referred to as the pressure matching. Assume that $N$ control points are placed over $\Omega$ and their positions are denoted by $\bm{r}_{\mathrm{c},n}$ ($n\in\{1,\ldots,N\}$). The cost function $J$ is approximated as the error between the synthesized and desired pressures at the control points. The optimization problem of pressure matching is written as
\begin{align}
 \minimize_{\bm{d}\in\mathbb{C}^L} \| \bm{Gd} - \bm{u}^{\mathrm{des}} \|^2 + \eta \| \bm{d} \|^2,
 \label{eq:opt_pm}
\end{align}
where $\bm{d} = [d_1, \ldots, d_L]^{\mathsf{T}} \in \mathbb{C}^L$  is the vector of the driving signals, $\bm{u}^{\mathrm{des}}=[u_{\mathrm{des}}(\bm{r}_{\mathrm{c},1}), \ldots, u_{\mathrm{des}}(\bm{r}_{\mathrm{c},N})]^{\mathsf{T}} \in \mathbb{C}^N$ is the vector of the desired sound pressures, and 
%$\bm{G}=[\bm{g}(\bm{r}_{\mathrm{c},1}), \ldots, \bm{g}(\bm{r}_{\mathrm{c},N})]^{\mathsf{T}} \in \mathbb{C}^{N \times L}$ 
$\bm{G} \in \mathbb{C}^{N \times L}$ is the matrix consisting of the transfer functions $g_l(\bm{r}_{\mathrm{c},n})$ between $L$ secondary sources and $N$ control points. The second term is the regularization term to prevent an excessively large amplitude of $\bm{d}$, and $\eta$ is a constant parameter. The closed-form solution of Eq.~\eqref{eq:opt_pm} is obtained as
\begin{align}
 \hat{\bm{d}} = \left( \bm{G}^{\mathsf{H}}\bm{G} + \eta \bm{I} \right)^{-1} \bm{G}^{\mathsf{H}} \bm{u}^{\mathrm{des}}.
\label{eq:drv_pm}
\end{align}

Another strategy to approximately solve Eq.~\eqref{eq:cost} is to represent the sound field by spherical wavefunction expansion~\cite{Williams:FourierAcoust,Martin:MultScat}, which is referred to as mode matching~\cite{Poletti:J_AES_2005}. The driving signal is obtained so that the expansion coefficients of the synthesized sound field coincide with those of the desired sound field. The mode matching method is generalized by introducing a weighting matrix consisting of the regional integration of the spherical wavefunctions, which is called weighted mode matching~\cite{Ueno:IEEE_ACM_J_ASLP2019}. Instead of empirical truncation of the expansion order in mode matching, optimal weights on the expansion coefficients are applied in weighted mode matching. 

\section{\uppercase{Weighted pressure matching}}

Since pressure matching is based on the discrete approximation of the target region, the region between the control points is taken into consideration. We consider incorporating a sound field interpolation method into pressure matching. First, we introduce the kernel interpolation method for sound fields. Then, the weighted pressure matching is formulated by representing the sound field using the kernel interpolation. 

\subsection{Kernel interpolation of sound field} 

The goal of the sound field interpolation is to estimate the pressure distribution $u(\bm{r})$ from the discrete set of pressure measurements $s_m$ at $\bm{r}_{m}$ ($m\in\{1,\ldots,M\}$). This interpolation problem is formulated as
\begin{align}
 \minimize_{u\in\mathcal{H}} \sum_{m=1}^M |u(\bm{r}_m) - s_m|^2 + \lambda \|u\|_{\mathcal{H}}^2,
\label{eq:opt_krr}
\end{align}
where $\mathcal{H}$ is a function space for which we seek a solution, $\|\cdot\|_{\mathcal{H}}$ is a norm on $\mathcal{H}$, and $\lambda$ is a constant parameter. When $\mathcal{H}$ is a function space called a \textit{reproducing kernel Hilbert space}, the optimization problem \eqref{eq:opt_krr} corresponds to kernel ridge regression, for which a closed-form solution can be obtained~\cite{Murphy:ML}. Here, $\mathcal{H}$ is assumed to be a reproducing kernel Hilbert space with the inner product $\langle \cdot, \cdot \rangle_{\mathcal{H}}$ and the positive-definite reproducing kernel $\kappa : \mathcal{H} \times \mathcal{H} \to \mathbb{C}$. On the basis of the representer theorem~\cite{Scholkopf:COLT2001}, the solution for Eq.~\eqref{eq:opt_krr} is obtained as
\begin{align}
 u(\bm{r}) = \bm{\kappa}(\bm{r})^{\mathsf{T}} \left( \bm{K} + \lambda \bm{I} \right)^{-1} \bm{s},
\label{eq:ker_interp}
\end{align}
where
\begin{align}
 \bm{\kappa}(\bm{r}) &= 
\begin{bmatrix}
 \kappa(\bm{r},\bm{r}_1) & \ldots & \kappa(\bm{r},\bm{r}_M)
\end{bmatrix}^{\mathsf{T}}\\
\bm{K} &= 
\begin{bmatrix}
 \kappa(\bm{r}_1,\bm{r}_1) & \cdots & \kappa(\bm{r}_1,\bm{r}_M) \\
 \vdots & \ddots & \vdots \\
 \kappa(\bm{r}_M,\bm{r}_1) & \cdots & \kappa(\bm{r}_M,\bm{r}_M)
\end{bmatrix}\\
\bm{s} &= 
\begin{bmatrix}
 s_1 & \ldots & s_M
\end{bmatrix}^{\mathsf{T}}.
\end{align}

Next, it is necessary to define appropriate $\mathcal{H}$ and $\langle \cdot , \cdot \rangle_{\mathcal{H}}$, as well as $\kappa$. The pressure field $u$ inside the source-free simply connected region $\Omega$ can be modeled as a solution of the homogeneous Helmholtz equation as
\begin{align}
 (\nabla^2 + k^2) u = 0,
\label{eq:helm}
\end{align}
where $\nabla^2$ is the Laplacian and $k=\omega/c$ is the wavenumber defined with the sound velocity $c$. Any solution of Eq.~\eqref{eq:helm} can be well approximated by the superposition of plane waves, i.e., the Herglotz wavefunction~\cite{Colton:InvAcoust_2013,Ueno:IEEE_J_SP_2021}, as
\begin{align}
 \hat{u}(\bm{r}) = \frac{1}{4\pi} \int_{\mathbb{S}_2} \tilde{u}(\bm{\xi}) \mathrm{e}^{-\mathrm{j}\bm{k}^{\mathsf{T}}\bm{r}} \mathrm{d}\bm{\xi},
\end{align}
where $\mathbb{S}_2$ is the unit sphere, $\tilde{u}(\bm{\xi})$ is the (square-integrable) complex amplitude of the plane wave of arrival direction $\bm{\xi}\in\mathbb{S}_2$, and $\bm{k}=-k\bm{\xi}$ is the wave vector. Using this representation, we define the inner product and norm over the Hilbert space $\mathcal{H}$ as
\begin{align}
 \langle u_1, u_2 \rangle_{\mathcal{H}} &= \frac{1}{4\pi} \int_{\mathbb{S}_2} \frac{1}{\gamma(\bm{\xi})} \tilde{u}_1(\bm{\xi})^{\ast} \tilde{u}_2(\bm{\xi}) \mathrm{d}\bm{\xi}, \\
\|u\|_{\mathcal{H}} &= \sqrt{ \langle u, u \rangle_{\mathcal{H}} },
\end{align}
respectively. Here, $\gamma(\bm{\xi})$ is a directional weighting function, which is introduced to incorporate prior knowledge on source directions. We set the kernel function $\kappa(\bm{r}_1,\bm{r}_2)$ as
\begin{align}
 \kappa(\bm{r}_1,\bm{r}_2) = \frac{1}{4\pi} \int_{\mathbb{S}_2} \gamma(\bm{\xi}) \mathrm{e}^{-\mathrm{j}\bm{k}^{\mathsf{T}}(\bm{r}_1-\bm{r_2})} \mathrm{d} \bm{\xi}.
\label{eq:kernel}
\end{align}
It can be confirmed that $\kappa(\bm{r}_1,\bm{r}_2)$ is the reproducing kernel of $\mathcal{H}$ because the inner product of $\kappa(\bm{r},\bm{r}^{\prime})$ and $u(\bm{r})$ equals $u(\bm{r}^{\prime})$. 

We define the directional weighting function as
\begin{align}
 \gamma(\bm{\xi}) = \mathrm{e}^{\rho \bm{\xi}^{\mathsf{T}} \hat{\bm{r}}},
\label{eq:dir_weight}
\end{align}
where $\rho \ge 0$ is a constant parameter and $\hat{\bm{r}}\in\mathbb{S}_2$ represents the prior arrival direction of the source. This function is derived from the von Mises--Fisher distribution in directional statistics~\cite{Mardia:DirStat}. For $\rho>0$, one can find that the smaller the $\rho \bm{\xi}^{\mathsf{T}}\hat{\bm{r}}$ is, the larger the norm $\|u\|_{\mathcal{H}}$ becomes, and vice versa. Therefore, the regularization term in Eq.~\eqref{eq:opt_krr} becomes larger when the difference between the prior arrival direction $\hat{\bm{r}}$ and the direction of $\bm{\xi}$ becomes larger. When $\rho=0$, the weighting function becomes uniform ($\gamma(\bm{\xi})=1$). The shape of $\gamma(\bm{\xi})$ becomes sharper with increasing $\rho$. 

By substituting Eq.~\eqref{eq:dir_weight} into Eq.~\eqref{eq:kernel}, we can derive the kernel function with directional weighting as
\begin{align}
 \kappa(\bm{r}_1,\bm{r}_2) &= 
\frac{1}{4\pi} \int_{\mathbb{S}_2} \mathrm{e}^{\rho \bm{\xi}^{\mathsf{T}} \hat{\bm{r}}} \cdot \mathrm{e}^{-\mathrm{j}\bm{k}^{\mathsf{T}}(\bm{r}_1-\bm{r_2})} \mathrm{d} \bm{\xi} \notag\\
&= j_0\left( \left[ \left( \mathrm{j}\rho \sin\theta\cos\phi - k x_{12} \right)^2 + \left(\mathrm{j}\rho \sin\theta\sin\phi - ky_{12}\right)^2 + \left(\mathrm{j}\rho\cos\theta - kz_{12} \right)^2 \right]^{\frac{1}{2}} \right),
\label{eq:kernel_dir}
\end{align}
where $j_0(\cdot)$ is the 0th-order spherical Bessel function of the first kind, $\phi$ and $\theta$ are the azimuth and zenith angles of $\hat{\bm{r}}$, and $\bm{r}_1 - \bm{r}_2 = [x_{12}, y_{12}, z_{12}]^{\mathsf{T}}$. By setting $\rho=0$, we can simplify Eq.~\eqref{eq:kernel_dir}, and obtain the kernel function of uniform weighting as 
\begin{align}
 \kappa(\bm{r}_1,\bm{r}_2) &= \frac{1}{4\pi} \int_{\mathbb{S}_2} \mathrm{e}^{-\mathrm{j}\bm{k}^{\mathsf{T}}(\bm{r}_1-\bm{r_2})} \mathrm{d} \bm{\xi} \notag\\
&= j_0(k \|\bm{r}_1 - \bm{r}_2 \|).
\label{eq:kernel_uni}
\end{align}
Thus, the kernel interpolation of a sound field is achieved by using Eq.~\eqref{eq:ker_interp} with the kernel function in Eq.~\eqref{eq:kernel_dir} or \eqref{eq:kernel_uni}. These kernel functions are derived for a three-dimensional (3D) sound field, but similar kernel functions can be derived for a two-dimensional (2D) sound field~\cite{Koyama:IEEE_ACM_J_ASLP2021}.

\subsection{Weighted pressure matching based on kernel interpolation} 

We apply the kernel interpolation in the sound field reproduction to accurately approximate the cost function $J$ in Eq.~\eqref{eq:cost} from the pressures at the control points. A similar idea has been applied in the context of spatial active noise control~\cite{Ito:ICASSP2019,Koyama:IEEE_ACM_J_ASLP2021,Arikawa:ICASSP2022} and multizone sound field control~\cite{Brunnstroem:ICASSP2022}. The transfer functions $g_l(\bm{r})$ and desired sound field $u_{\mathrm{des}}(\bm{r})$ are interpolated from those at the control points as
\begin{align}
 \hat{g}_l(\bm{r}) &= \bm{\kappa}_l(\bm{r})^{\mathsf{T}} \left( \bm{K}_l + \lambda \bm{I} \right)^{-1} \bm{g}_l := \bm{z}_l(\bm{r})^{\mathsf{T}} \bm{g}_l\\
 \hat{u}_{\mathrm{des}} &= \bm{\kappa}^{\mathrm{des}}(\bm{r})^{\mathsf{T}} \left( \bm{K}^{\mathrm{des}} + \lambda \bm{I} \right)^{-1} \bm{u}^{\mathrm{des}}:= \bm{z}^{\mathrm{des}}(\bm{r})^{\mathsf{T}} \bm{u}^{\mathrm{des}}, 
\end{align}
where $\bm{g}_l\in\mathbb{C}^L$ is the $l$th column vector of $\bm{G}$, and $\bm{\kappa}_l(\bm{r}),\bm{\kappa}^{\mathrm{des}}(\bm{r})\in\mathbb{C}^N$ and $\bm{K}_l,\bm{K}^{\mathrm{des}}\in\mathbb{C}^{N \times N}$ are respectively the vectors and matrices consisting of the kernel function defined with the positions $\{\bm{r}_{\mathrm{c},n}\}_{n=1}^N$. Then, the cost function $J$ can be approximated using $\hat{\bm{g}}(\bm{r})=[\bm{z}_1(\bm{r})^{\mathsf{T}}\bm{g}_1, \ldots, \bm{z}_L(\bm{r})^{\mathsf{T}}\bm{g}_L]^{\mathsf{T}}$ as
\begin{align}
 J &\approx \int_{\Omega} \left| \sum_{l=1}^L d_l \hat{g}_l(\bm{r}) - \hat{u}_{\mathrm{des}}(\bm{r}) \right|^2 \mathrm{d}\bm{r} \notag\\
 %&= \int_{\Omega} \left| \sum_{l=1}^L d_l \bm{z}_l(\bm{r})^{\mathsf{T}} \bm{g}_l - \bm{z}^{\mathrm{des}}(\bm{r})^{\mathsf{T}} \bm{u}^{\mathrm{des}} \right|^2 \mathrm{d}\bm{r} \notag\\
 &= \int_{\Omega} \left| \hat{\bm{g}}(\bm{r})^{\mathsf{T}}\bm{d} - \bm{z}^{\mathrm{des}}(\bm{r})^{\mathsf{T}}\bm{u}^{\mathrm{des}} \right|^2 \mathrm{d}\bm{r} \notag\\
&= \bm{d}^{\mathsf{H}} \bm{W}_{gg} \bm{d} - \bm{d}^{\mathsf{H}} \bm{W}_{gu} \bm{u}^{\mathrm{des}} + C,
\label{eq:cost_wmm}
\end{align}
where 
\begin{align}
 \bm{W}_{gg} &= \int_{\Omega} \hat{\bm{g}}(\bm{r})^{\ast} \bm{g}(\bm{r})^{\mathsf{T}} \mathrm{d}\bm{r} \\
 \bm{W}_{gu} &= \int_{\Omega} \hat{\bm{g}}(\bm{r})^{\ast} \bm{z}^{\mathrm{des}}(\bm{r}^{\mathsf{T}}) \mathrm{d}\bm{r},
\end{align}
and $C$ is the term not including $\bm{d}$. Therefore, the optimal driving signal $\bm{d}$ is obtained by solving
\begin{align}
 \minimize_{\bm{d}\in\mathbb{C}^L} \bm{d}^{\mathsf{H}} \bm{W}_{gg} \bm{d} - \bm{d}^{\mathsf{H}} \bm{W}_{gu} \bm{u}^{\mathrm{des}} + \eta \|\bm{d}\|^2.
\label{eq:opt_wpm}
\end{align}
Again, the regularization term is added. This minimization problem also has the closed-form solution as
\begin{align}
 \hat{\bm{d}} = \left( \bm{W}_{gg} + \eta \bm{I} \right)^{-1} \bm{W}_{gu} \bm{u}^{\mathrm{des}}.
\label{eq:drv_wpm}
\end{align}
Since the directions of the secondary sources are generally known, the kernel functions for interpolating $g_l(\bm{r})$ as well as $\bm{z}_l(\bm{r})$ can be defined in advance by using Eq.~\eqref{eq:kernel_dir}. When the desired sound field is set with the parameters of source directions and positions, $\hat{\bm{r}}$ for $u_{\mathrm{des}}(\bm{r})$ can also be given. When it is difficult to set $\hat{\bm{r}}$ in advance, the kernel function of uniform weighting Eq.~\eqref{eq:kernel_uni} will be appropriate for defining $\bm{z}_l(\bm{r})$ and/or $\bm{z}^{\mathrm{des}}(\bm{r})$.

When using the same kernel function for interpolating $g_l(\bm{r})$ and $u_{\mathrm{des}}(\bm{r})$, e.g., the kernel function of uniform weighting Eq.~\eqref{eq:kernel_uni}, the cost function $J$ in \eqref{eq:cost_wmm} can be further simplified. By defining $\bm{z}(\bm{r})^{\mathsf{T}}=\bm{\kappa}(\bm{r})^{\mathsf{T}}(\bm{K}+\lambda\bm{I})^{-1}$, we can obtain  $J$ as
\begin{align}
 J &\approx \int_{\Omega} \left| \bm{z}(\bm{r})^{\mathsf{T}} \left(\bm{Gd} - \bm{u}^{\mathrm{des}} \right) \right|^2 \notag\\
 &= \left( \bm{Gd} - \bm{u}^{\mathrm{des}} \right)^{\mathsf{H}} \bm{W} \left( \bm{Gd} - \bm{u}^{\mathrm{des}} \right),
\label{eq:cost_wmm_s}
\end{align}
where
\begin{align}
 \bm{W} = \int_{\Omega} \bm{z}(\bm{r})^{\ast} \bm{z}(\bm{r})^{\mathsf{T}} \mathrm{d} \bm{r}.
\end{align}
Thus, the optimal driving signal is derived as
\begin{align}
 \hat{\bm{d}} = \left( \bm{G}^{\mathsf{H}}\bm{W}\bm{G} + \eta \bm{I} \right)^{-1} \bm{G}^{\mathsf{H}}\bm{W} \bm{u}^{\mathrm{des}}.
\label{eq:drv_wpm_s}
\end{align}
This driving signal can be regarded as the solution of the weighted mean square error between synthesized and desired pressures at the control points. Therefore, we refer to the proposed method as weighted pressure matching. The weighted pressure matching enables the increase in the reproduction accuracy of pressure matching only by introducing the weighting matrix. Note that the matrices $\bm{W}$, $\bm{W}_{gg}$, and $\bm{W}_{gu}$ can be computed only with the positions of the control points and the target region $\Omega$ by defining the kernel function using Eq.~\eqref{eq:kernel_dir} and/or \eqref{eq:kernel_uni}.

\section{\uppercase{Experiments}}

\begin{figure}
\centering
\includegraphics[width=6cm,clip]{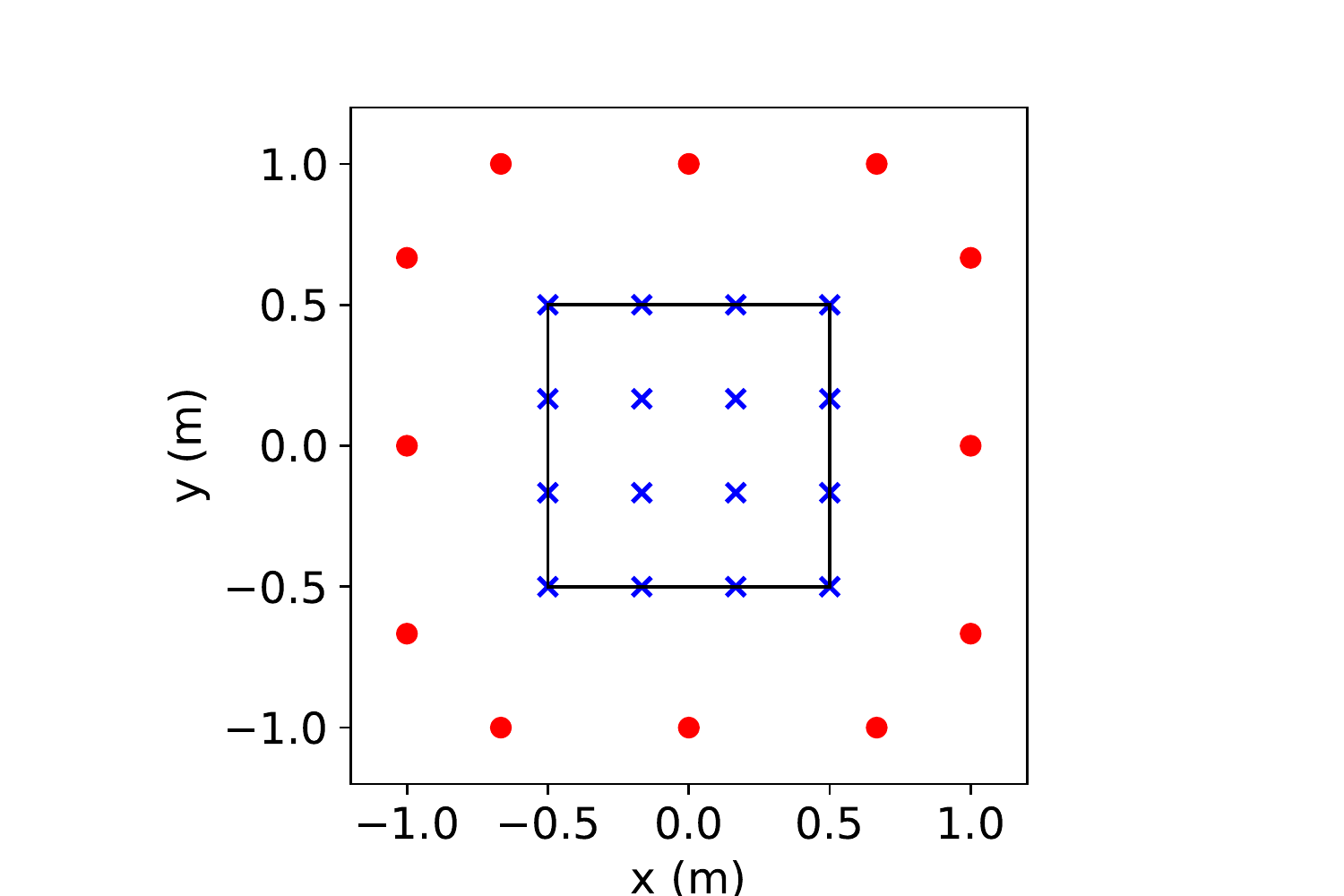}
\caption{Experimental setup. The target region was set as a 2D square region. Red dots and blue crosses indicate loudspeakers and control points, respectively.}
\label{fig:exp_cond}
\end{figure}

We conducted numerical experiments to evaluate the proposed method. Pressure matching and weighted pressure matching are hereafter denoted as PM and WPM, respectively. We also evaluated WPM with directional weighting, which is hereafter denoted as WPM (directional). In the experiments, we assumed a 2D free field. 

As shown in Fig.~\ref{fig:exp_cond}, 12 loudspeakers were placed at the same intervals along the border of a square with dimensions of $2.0~\mathrm{m} \times 2.0~\mathrm{m}$. The target region $\Omega$ was set as a square region of $1.0~\mathrm{m} \times 1.0~\mathrm{m}$. The centers of these squares were at the origin. Sixteen  control points were placed at the same intervals over the target region. In Fig.~\ref{fig:exp_cond}, the loudspeakers and control points are indicated by red dots and blue crosses, respectively. Each loudspeaker was assumed to be a point source. The desired sound field was set to be a single plane-wave field, whose propagation direction was $\pi/4~\mathrm{rad}$.

$\bm{u}^{\mathrm{des}}$ and $\bm{G}$ were given as pressure measurements at the control points. In WPM, the kernel function was set to be uniform, i.e., Eq.~\eqref{eq:kernel_uni}. The kernel function in WPM (directional) was the directional kernel given by Eq.~\eqref{eq:kernel_dir}, where the direction $\hat{\bm{r}}$ was set as the true directions of primary and secondary sources, and the parameter $\rho$ was $5.0$. The regularization parameter $\lambda$ in Eq.~\eqref{eq:ker_interp} and $\eta$ in Eqs.~\eqref{eq:drv_pm}, \eqref{eq:drv_wpm}, and \eqref{eq:drv_wpm_s} were set as $10^{-6}$. For evaluation measure, we define the signal-to-distortion ratio (SDR) as
\begin{align}
\mathrm{SDR}(\omega) = \frac{\int_{\Omega} |u_{\mathrm{des}}(\bm{r},\omega)|^2 \mathrm{d}\bm{r}}{\int_{\Omega} |u_{\mathrm{syn}}(\bm{r},\omega) - u_{\mathrm{des}}(\bm{r},\omega)|^2 \mathrm{d}\bm{r}},
\end{align}
where the integration was computed at the evaluation points regularly distributed over the target region. The intervals of the evaluation points were $10^{-2}~\mathrm{m}$.

\begin{figure}
\centering
\includegraphics[width=11cm,clip]{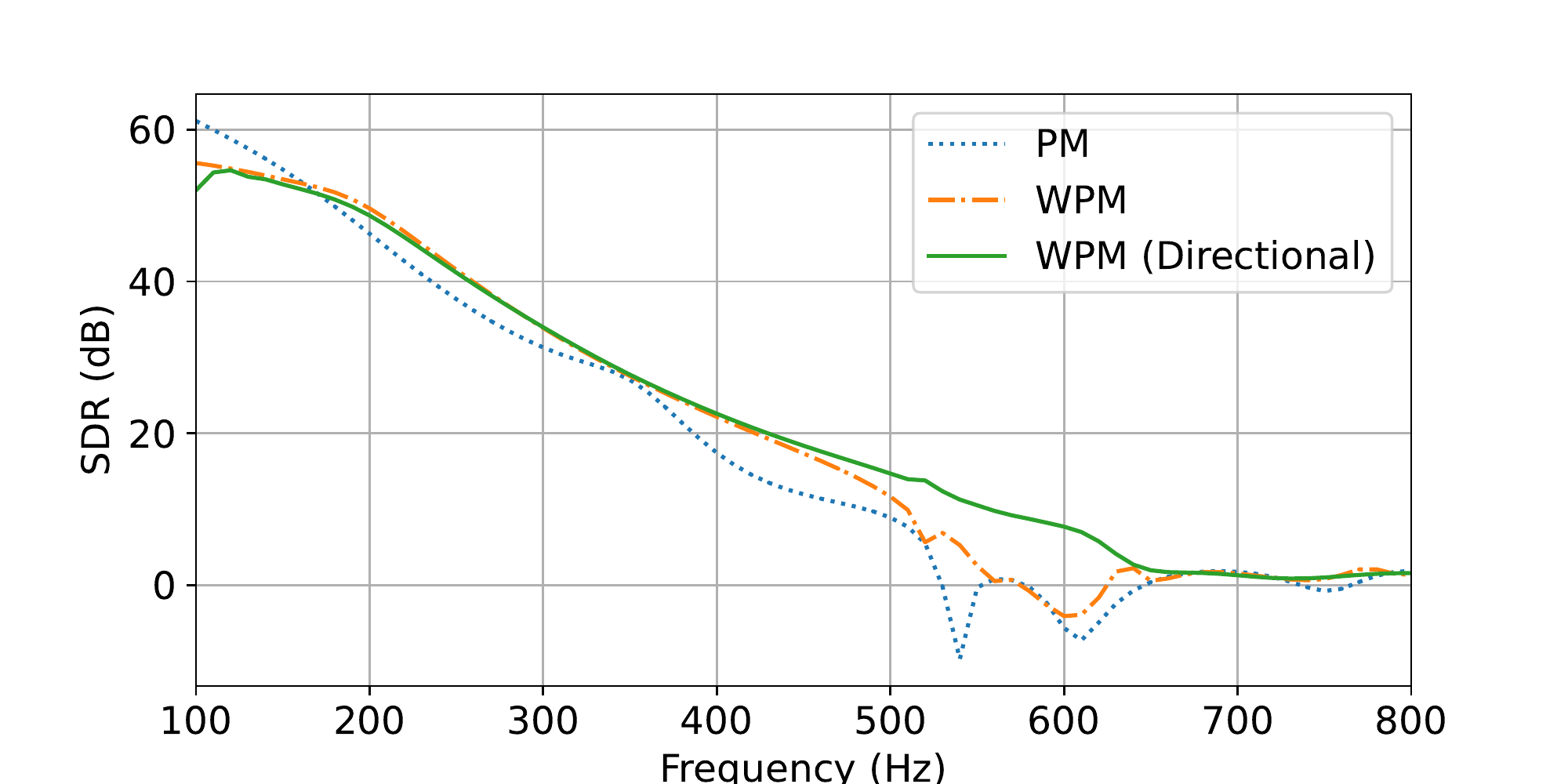}
\caption{SDR with respect to frequency.}
\label{fig:sdr}
%\end{figure}
%\begin{figure}
\centering
\subfloat[PM]{\includegraphics[width=5.5cm,clip]{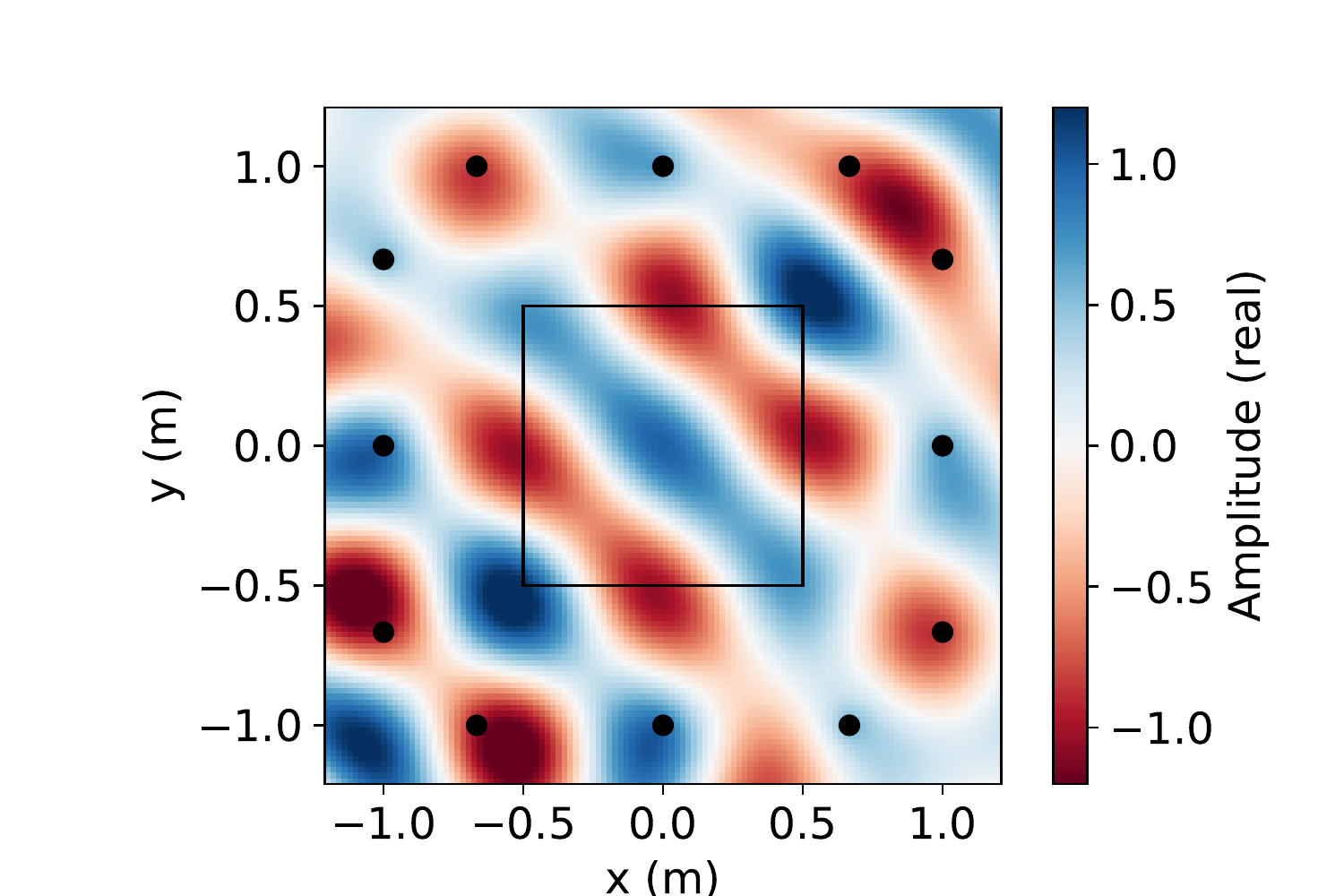}}
\subfloat[WPM]{\includegraphics[width=5.5cm,clip]{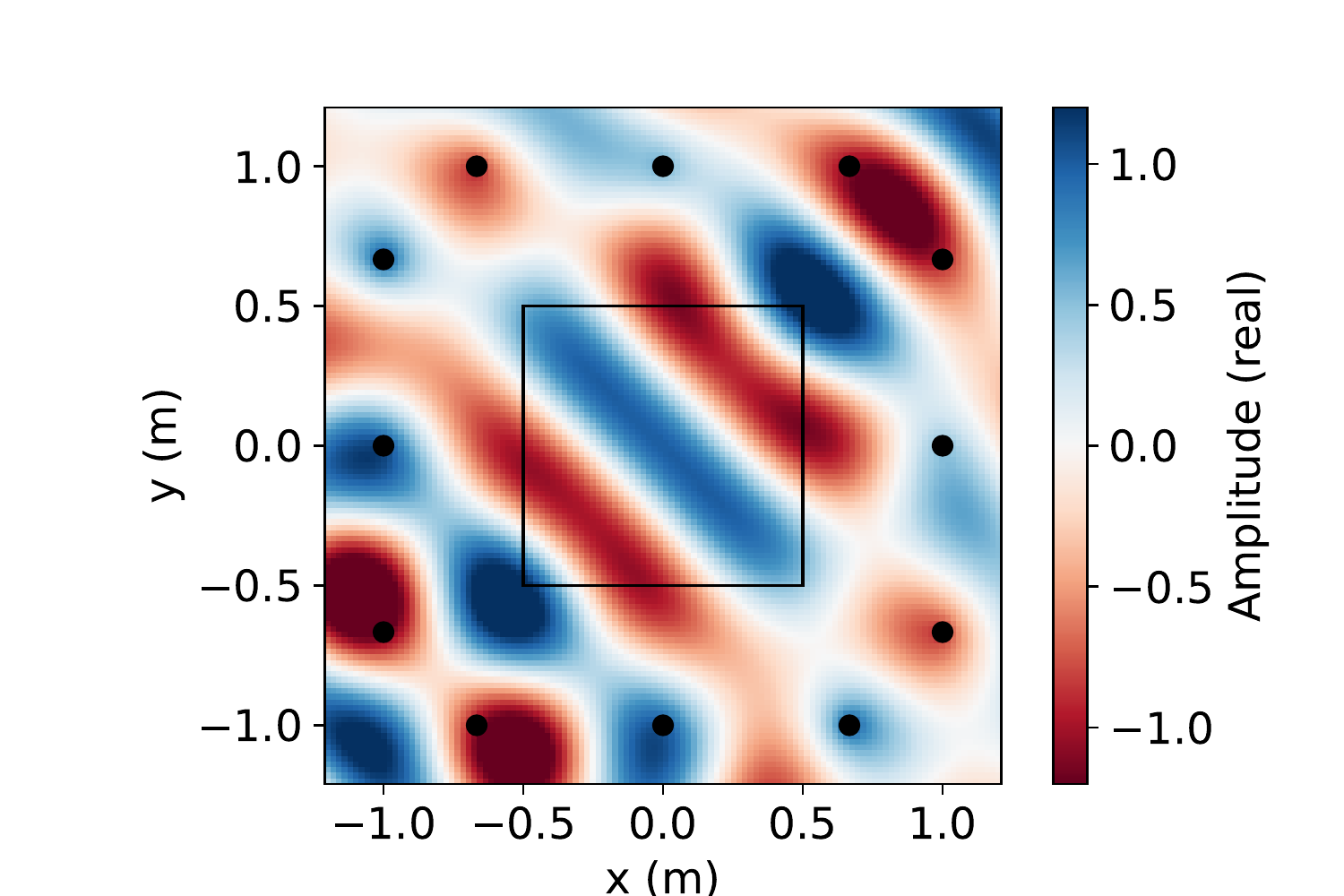}}
 \subfloat[WPM (Directional)]{\includegraphics[width=5.5cm,clip]{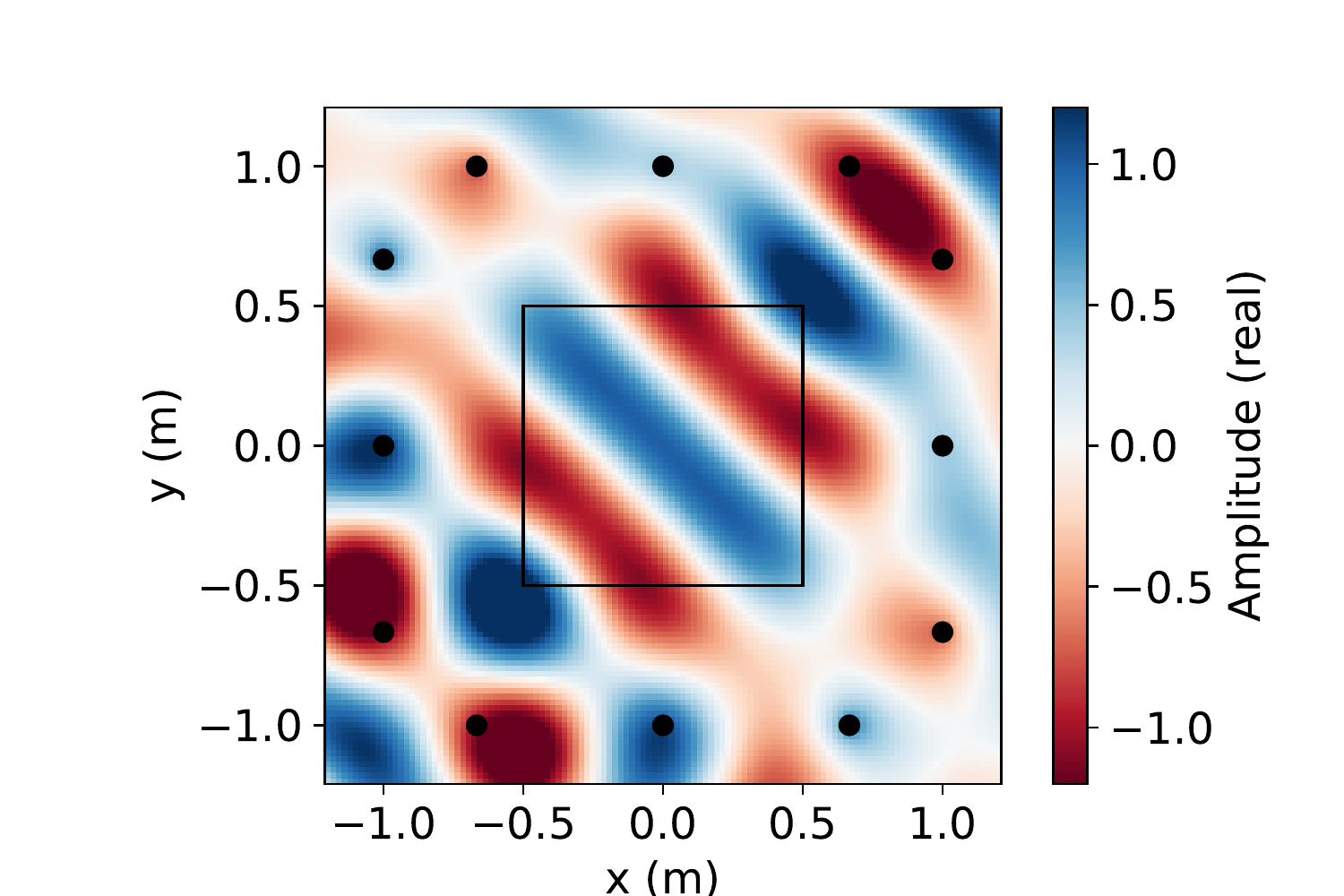}}
\caption{Synthesized pressure distributions at $450~\mathrm{Hz}$.}
\label{fig:exp_dist}
%\end{figure}
%\begin{figure}
\centering
\subfloat[PM]{\includegraphics[width=5.5cm,clip]{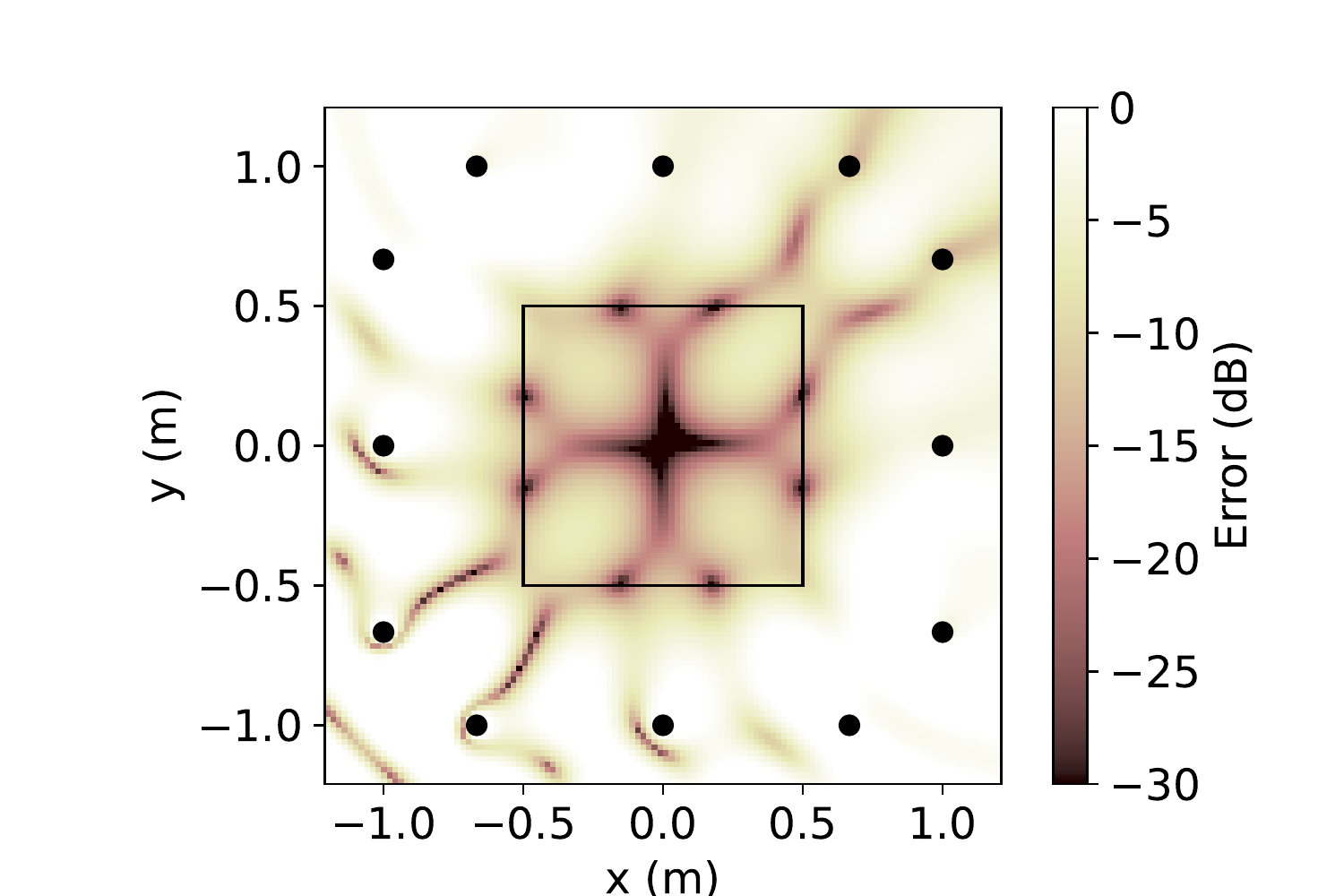}}
\subfloat[WPM]{\includegraphics[width=5.5cm,clip]{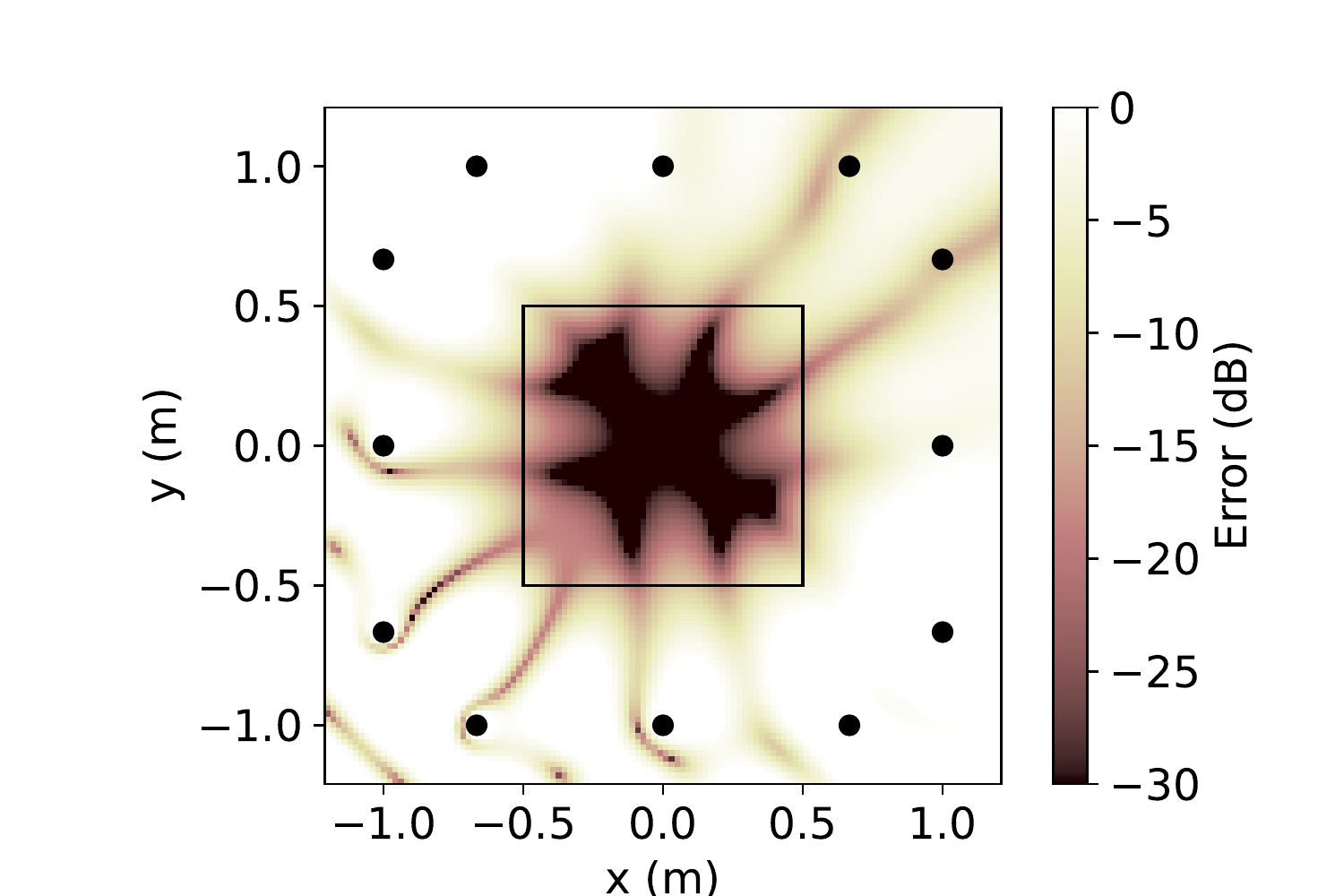}}
 \subfloat[WPM (Directional)]{\includegraphics[width=5.5cm,clip]{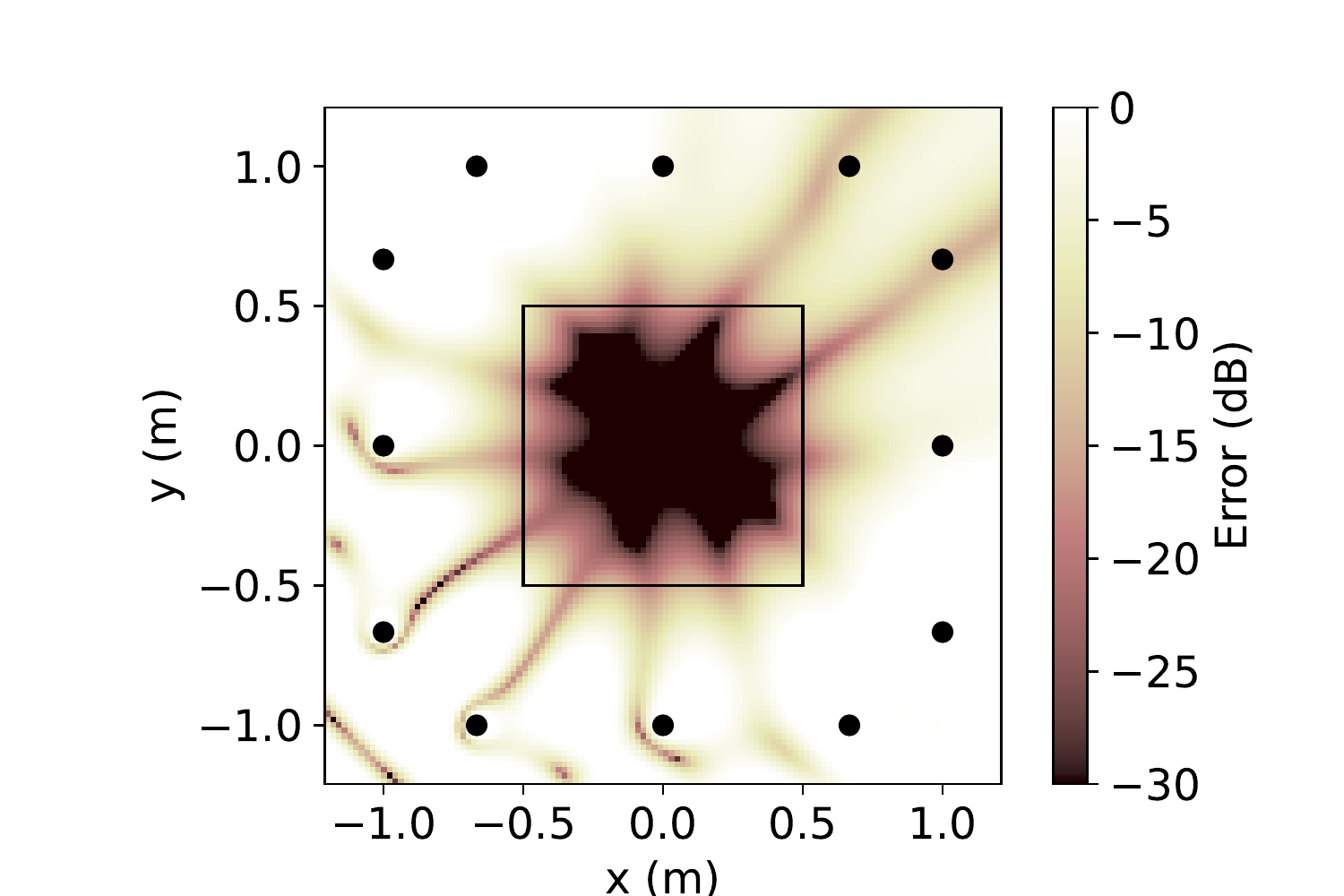}}
\caption{Square error distributions at $450~\mathrm{Hz}$. The SDRs of PM, WPM, and WPM (directional) were $11.9$, $17.3$, and $18.3~\mathrm{dB}$, respectively.}
\label{fig:exp_err}
\end{figure}

The SDR with respect to the frequency is plotted in Fig.~\ref{fig:sdr}. The SDRs were larger than $20~\mathrm{dB}$ at frequencies below $390~\mathrm{Hz}$. Above $400~\mathrm{Hz}$, the SDRs of WPM and WPM (directional) were larger than that of PM. In particular, WPM (directional) maintained a large SDR at high frequencies. In Figs.~\ref{fig:exp_dist} and \ref{fig:exp_err}, the synthesized pressure distribution and square error distribution at $450~\mathrm{Hz}$ of each method are shown. The region of small errors in WPM was larger than that of PM. The SDRs of PM, WPM, and WPM (directional) were $11.9$, $17.3$, and $18.3~\mathrm{dB}$, respectively. It can be considered that these differences are the effect of the interpolation. By taking into consideration the region between the control points, we can improve the reproduction accuracy. Since the interpolation accuracy of the directional kernel is high because of the use of prior information on the source directions, WPM (directional) achieved the highest reproduction accuracy. 

\section{\uppercase{Conclusion}}

We proposed a sound field reproduction method called weighted pressure matching. Pressure matching is a widely used optimization-based sound field reproduction method because of its simplicity. Since pressure matching is based on the synthesis of desired pressures at a discrete set of control points distributed over the target region, the region between the control points is not taken into consideration. Our cost function is defined as the regional integration of the synthesis error over the target region. On the basis of the kernel interpolation of sound fields, the driving signal of weighted pressure matching is obtained as the weighted least squares solution with the weighting matrix consisting of the regional integration of the kernel functions. In the numerical experiments, the weighted pressure matching achieved high reproduction accuracy compared with conventional pressure matching, especially at high frequencies. By introducing the directional weighting for interpolating the sound field, we can further increase the reproduction accuracy. 

\section*{\uppercase{Acknowledgments}}
This work was supported by JST FOREST Program (Grant Number JPMJFR216M, Japan).

%\hl{AO: I have not updated the reference style yet...}

\bibliographystyle{abbrv}
\renewcommand{\refname}{\normalfont\selectfont\normalsize}
\noindent \section*{\uppercase{References}}
\vspace{-18pt}

\bibliography{str_def_abrv,koyama_en,refs}
%\begin{thebibliography}{5}

%\bibitem{ref1}
%Author, A.; Author, B. Paper title. Journal name, volume (number), year, pages.

%\end{thebibliography}

\end{document}